# COMBINED AND COMPARATIVE ANALYSIS OF POWER SPECTRA

P.A. Sturrock[1], J. D. Scargle[2], G.Walther[3], and M.S. Wheatland[4]

[1]Center for Space Science and Astrophysics, Stanford University, Stanford, CA 94305-4060
[2]NASA/Ames Research Center, MS 245-3, Moffett Field, CA 94035
[3]Statistics Department, Stanford University, Stanford, CA 94305-4065
[4]School of Physics, University of Sydney, Sydney, Australia

## ABSTRACT

In solar physics, especially in exploratory stages of research, it is often necessary to compare the power spectra of two or more time series. One may, for instance, wish to estimate what the power spectrum of the combined data sets might have been, or one may wish to estimate the significance of a particular peak that shows up in two or more power spectra. One may also on occasion need to search for a complex of peaks in a single power spectrum, such as a fundamental and one or more harmonics, or a fundamental plus sidebands, etc. Visual inspection can be revealing, but it can also be misleading. This leads one to look for one or more ways of forming statistics, which readily lend themselves to significance estimation, from two or more power spectra. We derive formulas for statistics formed from the sum, the minimum, and the product of two or more power spectra. A distinguishing feature of our formulae is that, if each power spectrum has an exponential distribution, each statistic also has an exponential distribution.

The statistic formed from the minimum power of two or more power spectra is well known and has an exponential distribution. The sum of two or more powers also has a well-known distribution that is not exponential, but a simple operation does lead to an exponential distribution. Concerning the product of two or more power spectra, we find an analytical expression for the case n = 2, and a procedure for computing the statistic for n >2. We also show that some quite simple expressions give surprisingly good approximations.

1. INTRODUCTION

Since the Sun has been observed more or less routinely for many years and in a variety of modes (sunspots, radio, UV, X-ray, irradiance, etc.), it is frequently necessary to compare two or more solar data sets. This need may come about in the analysis of some specific hypothesis. However, one may also carry out such comparisons in an exploratory phase of research in which one is inspecting new data, or in the early stage of theoretical study. One common and useful technique is to carry out correlation analysis of the time series. For instance, this approach has been used extensively in research concerning solar neutrinos, since one may search for evidence of variability of the neutrino flux by determining whether neutrino flux measurements are correlated with measurements of some other solar variable such as the Wolf sunspot number (Bahcall, Field, & Press 1987; Bahcall & Press 1991; Bieber et al. 1990; Dorman & Wolfendale 1991), the surface magnetic field strength (Massetti & Storini 1993; Oakley et al. 1994), the intensity of the green-line corona (Massetti & Storini 1996), or the solar wind flux (McNutt 1995).

However, one may be particularly interested in oscillations in solar variables due, for instance, to the solar cycle, to solar rotation, or to other oscillations such as that discovered by Rieger et al. (1984) and others studied by Bai (1992), Bai & Cliver (1990), and Wolff (1976). Focusing on oscillations has the advantage that it may make it possible to "pull a signal out of the noise." In this situation, it would be helpful to have one or more procedures for combining and comparing time series that focus on known types of oscillations. The same techniques may then prove useful in searching for new types of oscillations.

Power spectrum analysis is one of the most powerful tools for the study of time series. (See, for instance, Oppenheim, Schafer & Buck 1999.) In some cases, it may be possible to actually combine the time series and carry out a power spectrum analysis of the combined data. However, this is not always possible, and it may not be convenient even when it is possible. In such cases, it may be more convenient to compare and contrast the power spectra derived from the different time series.

In addition to looking for similarities between two different power spectra, one may need to determine whether there is a significant relationship between oscillations at two or more frequencies in the same power spectrum. For instance, we have found that the power spectrum formed from Homestake (Davis & Cox 1991; Lande et al. 1992; Cleveland et al. 1995, 1998) measurements shows not only a peak at 12.88 $y^{-1}$ due apparently to solar rotation, but also two sidebands close to 11.88 $y^{-1}$ and 13.88 $y^{-1}$ due apparently to the influence of the inclination to the ecliptic of the Sun's rotation axis (Sturrock, Walther, & Wheatland 1997; Sturrock, Walther, & Wheatland 1998).

For these reasons, it is important to determine how much information one can obtain from power spectra alone. For the purpose of this article, we therefore assume that the time-series from which the spectra were derived are unknown, and that the only relevant available information comprises the power spectra and the probability distribution functions of the powers for the null hypothesis that the time-series contain no systematic oscillations (the "baseline" power spectrum). In some cases for which the properties of the time-series are well known, the probability distribution function of the powers may be well determined. However, in analyzing observational data, the probability distribution function will probably not be known *ab initio*, and it will be necessary to determine, or at least estimate, this function from the data. We comment further on this point later in this section. For the situation outlined, we need to construct one or more procedures for determining whether two or more peaks in the same or different power spectra are related. It is desirable that these procedures should provide some kind of significance estimate for any such relationship.

It is well known that if a time series is formed from a sequence of measurements of a normally distributed random variable with mean zero and variance unity, the power S at any specified frequency $\nu$, which may be computed in this simple case as the "periodogram" (see, for instance, Bai 1992; Scargle 1982; Horne & Baliunas 1986; Schuster 1905), is distributed exponentially:

$$P(S)dS = e^{-S}dS. \tag{1.1}$$

In more complex and more realistic situations, this simple result will not be applicable. Indeed, Rayleigh was aware early on of some of the limitations of this method (Rayleigh 1903). Nevertheless, it is convenient for our purposes to focus on the case that the probability distribution function for the power is given by (1.1). In this case, the probability of obtaining a power S or more is given by the cumulative distribution function

$$C(S) = \int_S^\infty P(x)dx = e^{-S}. \tag{1.2}$$

Since S is a function of the frequency $\nu$, P and C also are expressible as functions of $\nu$.

In more realistic situations, one must expect that the power will be distributed in a more complex manner. However, one can in principle "normalize" the power spectrum, so that it is distributed exponentially, if the "baseline" distribution is known or can be estimated. One may be able to make this estimate from knowledge of the statistical properties of the time series, or one may be able to make a reasonable estimate of the probability distribution of the power empirically by inspection of the power spectrum over a wide frequency range. It may also be possible to use Monte Carlo simulation or a similar procedure for assessing the baseline power spectrum, in which case one need not assume this function to be frequency-independent. Estimation of the baseline distribution is an active area of research (see, for instance, Thomson 2000; Thomson, Lanzerotti, & Maclennan 2001). One should note that uncertainties in the initial power spectra propagate when power spectra are combined, and that this issue should be considered in any application. A general discussion of these issues lies outside the scope of this article.

We here assume that the power S is known to be distributed according to the probability density function $P(S)$ (that might be frequency-dependent) such that the probability that S lies in the range $S$ to $S + dS$ is given by $P(S)dS$. Then if we introduce the statistic S* defined by

$$S^* = -\ln(C) = -\ln\left(\int_S^\infty dS' P(S')\right), \qquad (1.3)$$

we see that S* is distributed exponentially (independent of frequency). In this sense, S* may be regarded as a "normalized" power. In the remainder of this article, we therefore restrict our attention to power estimates that are distributed exponentially. We could of course adopt C, for which the baseline is flat, as an alternative normalization of the power, but we prefer to use S* since it is familiar to time-series analysts, and since a logarithmic function provides a more convenient representation of extreme values.

From this point on, our goal is simply that of finding one or more procedures for estimating the significance of two or more related measurements of variables that have exponential distributions, selecting these procedures in such a way that the relevant statistics are also distributed exponentially. This brings a measure of uniformity to the analysis not unlike the uniformity in a market that follows the adoption of a common currency.

It is important to note that it is now irrelevant whether or not we are dealing with power spectra formed from time series. Studying the properties of combinations of variables that have exponential distributions is a purely mathematical exercise. This mathematical problem is reminiscent of a standard problem in statistics, that of estimating the significance of two or more related measurements of variables which, if random, have normal distributions. A familiar and very useful technique is that of computing the chi-square statistic. (See, for instance, Bartoszynki & Niewiadomska-Bugaj 1996, p. 757, or Rice 1988, p. 168.) When using chi-square statistics, it is necessary to consult tables to determine the significance level of a particular estimate, but we avoid this necessity by forming statistics that have exponential distributions. If the value of a statistic is found to be G, the significance level of this result is exactly the same as the significance level of a power spectrum that has the same value at a specified frequency. As a result, we may "daisy-chain" calculations. Suppose, for instance, that we have data for two different variables, each extending over several solar cycles, and we wish to investigate the possibility that each variable contains oscillations that show up in two or more solar cycles, and that some of these oscillations show up in both data sets. For each variable, we could regard data from different solar cycles as being independent, and then derive a measure of the combined evidence from power spectra formed from these solar cycles. Since each measure will have an exponential distribution, we can then compare the measures so derived for the two variables in the same way as if each were a simple power spectrum.

We can also apply the same procedures to combine and/or compare oscillations at two or more different but related frequencies in the same power spectrum. For instance, a power spectrum may contain a peak near the solar rotation frequency and peaks near harmonics of that frequency. By using the known theoretical relationship between a set of frequencies, we may use our techniques to combine or compare measurements that refer to different frequencies. However, one must be very careful: if we are combining or comparing peaks in power spectra derived from independent data sets, we are entitled to regard these peaks as independent variables, but if we are combining two or more power measurements from the same power spectrum, it may or may not be permissible to regard them as independent variables. For instance, if the data set comprises measurements taken at regular intervals $\Delta t$, the power spectrum will be contaminated by "aliasing," such that an oscillation at frequency $\nu$ will show up in the power spectrum not only at frequency $\nu$, but also at frequencies $\nu \pm \nu_T$, where $\nu_T = 1/\Delta t$. This contamination constitutes a severe problem in the analysis of solar neutrino data, since radiochemical experiments usually operate with runs that are separated by integral numbers of weeks, and data from Cerenkov experiments is typically presented in bins with highly regular timing (see, for instance, Sturrock 2004).

In order to make this article more readable and more useful, mathematical derivations are relegated to appendices. In Section 2, we introduce a statistic that is conceptually close to the chi-square statistic, first forming the sum of the variables (analogous to forming the sum of the squares of the variables in the context of normal distributions), and then applying a transformation such that the resulting distribution has an exponential distribution; we refer to this quantity as the "combined power statistic." In Section 3, we develop the "minimum power statistic," that is formed from the minimum (for each frequency value) of two or more power spectra. In Section 4, we introduce the "joint power statistic," that is formed from the product of two or more power spectra. In Section 5, we give the results of the application of these statistics to some simulated problems. Section 6 contains a brief discussion of the results of this article.

## 2. COMBINED POWER STATISTIC

If we wish to combine information from n independent power spectra, the combination that would correspond to the chi-square statistic is the sum of the powers, which we write as

$$Z = S_1 + S_2 + ... + S_n. \qquad (2.1)$$

It is well known that Z satisfies the gamma distribution with parameters n and 1. [See, for instance, Bickel & Doksum (1977), p. 13.]

We show in Appendix B, which uses some general results derived in Appendix A, that the following function of Z, which we refer to as the "combined power statistic," is distributed exponentially:

$$G_n(Z) = Z - \ln\left(1 + Z + \tfrac{1}{2}Z^2 + ... + \tfrac{1}{(n-1)!}Z^{n-1}\right). \qquad (2.2)$$

Figure 1 gives plots of the combined power statistics of orders 2, 3, and 4.

## 3. MINIMUM POWER STATISTIC

We may wish to determine the frequency for which the minimum power among two or more power spectra has the maximum value. We therefore consider the following quantity, formed from the independent variables $x_1$, $x_2$, …$x_n$, each of which is distributed exponentially:

$$U(x_1, x_2, ..., x_n) = Min(x_1, x_2, ..., x_n). \qquad (3.1)$$

It is well known that the following function of U, that we refer to as the "minimum power statistic," is distributed exponentially:

$$K_n(U) = nU. \qquad (3.2)$$

[See, for instance, Bickel & Doksum (1977), p. 46. It should also be noted that there can be problems in using the minimum power statistic: for instance, the false-alarm rate can be too high (Thomson 1977).]

## 4. JOINT POWER STATISTIC

We next consider forming something resembling a "correlation function" by forming the product of two or more independent power spectra. We first consider just two power spectra, since this case lends itself to analytical treatment. It proves convenient to work with the square root of the product (the geometric mean), and we therefore write

$$X = (S_1 S_2)^{1/2}. \qquad (4.1)$$

We show in Appendix C that the following function of X is distributed exponentially:

$$J_2 = -\ln(2X\, K_1(2X)), \qquad (4.2)$$

where $K_1$ is the Bessel function of the second kind.

We show in Appendix E that we may determine simple functional fits to the joint power statistics. The fit to $J_2$ is found to be

$$J_{2A} = \frac{1.943 X^2}{0.650 + X}. \qquad (4.3)$$

We show J2 and J2A in Figure 2. The difference (mean and standard deviation) between the true value and the approximate value is only $0.019 \pm 0.036$ for 21 evaluations between $X = 0$ and $X = 10$. This difference is fortunately negligible, so there is no need to make calculations via Equation (4.2) or to consult a table.

We now consider joint power statistics of higher orders, and consider the following geometric mean of n powers:

$$X = (S_1...S_n)^{1/n} . \qquad (4.4)$$

For n > 2, we have not succeeded in finding useful analytical functions of X that are distributed exponentially. The referee has kindly pointed out that this problem has been addressed by Lomnicki (1967) who shows that the probability distribution function of the product may be obtained by an operation on the sum of an infinite series of terms, each term involving an array of Euler psi functions and their derivatives. As an alternative to the Lomnicki procedures, we give in Appendix D a sequential procedure for computing these functions. Using the procedure of Appendix E, we find that the joint power statistics of third and fourth orders are given to very good approximation by the following expressions:

$$J_{3A} = \frac{2.916 X^2}{1.022 + X} , \qquad (4.5)$$

and

$$J_{4A} = \frac{3.881 X^2}{1.269 + X} . \qquad (4.6)$$

These approximate expressions are compared with the calculated values in Figure 2. For 21 evaluations of X in the range 0 to 10, $J_{3A} - J_3 = 0.014 \pm 0.028$, with extrema –0.07 and 0.02, and $J_{4A} - J_4 = -0.002 \pm 0.020$, with extrema –0.03 and 0.05. We see that the approximate expressions are sufficiently accurate that there is no need to make further calculations or to consult tables.

## 5. SIMULATIONS

It is interesting to calculate the preceding statistics for simple numerical examples. By analogy with our power spectrum studies of the solar neutrino flux, we have formed synthetic spectra for frequencies in the range 0.01 to 40, in steps of 0.01. (In our neutrino studies, these would be frequencies measured in cycles per year.) For each frequency, a power $S_1$ was chosen from an exponential distribution using $S_1 = -\ln(u)$, where u is a uniformly distributed random variable in the range $0 < u < 1$. For the chosen signal frequency $v_S = 20$, the power was set to a fixed value $P_S$. This procedure was repeated to produce the four spectra $S_1(v), ..., S_4(v)$. For the case $P_S = 5$, the spectra are shown in Figure 3. For this set of simulations, we have computed the three power statistics introduced in Sections 3, 4, and 5. A typical set of results is shown in Figure 4, where we see that the combined power statistic has the value 12.7 at the signal frequency [panel (a)]; the minimum power statistic has the value 20, as expected [panel (b)]; and the joint power statistic has the value 15.5 [panel (c)]. We find that it is typically the case that if four spectra have similar values at the signal frequency, the minimum power statistic has the largest value and the combined power statistic the smallest value. However, this ordering does not hold for an arbitrary set of powers.

As a second numerical test of the statistics, we have considered the application to spectra produced from synthetic time series. For a sequence of $N$ unit-spaced times $t_i = i$ ($i = 0,1,...,N-1$), four time series

$$X_j(t_i) = A\sin(2\pi v_0 t_i) + R_{j,i} \qquad (5.1)$$

were generated ($j = 1,2,3,4$), with a signal frequency $v_0 = 0.25$ per unit time, the noise term $R_{j,i}$ being generated from a normal distribution with zero mean and unit standard deviation. For each of these time series the classical periodogram (e.g. Scargle 1982)

$$S_j(\nu) = \frac{1}{N}\left|\sum_{k=0}^{N} X_j(t_k)\exp(-2\pi i k \nu)\right|^2 \qquad (5.2)$$

was evaluated at the usual set of $N/2+1$ frequencies $\nu = \nu_l = l/N$ ($l = 0,1,2,...,N/2$), using the Fast Fourier Transform. Note that for the choice of unit spacing in the time series the highest frequency in this set (the Nyqvist frequency) is 0.5 per unit time, so the chosen signal frequency is half the Nyqvist frequency. The mean power at the signal frequency in the periodograms is $\overline{S}_0 = N(A/2)^2 + 1$ (Scargle 1982). We have chosen $\overline{S}_0 = 5$ and $N = 8192$, so that $A = 2\left[(\overline{S}_0 - 1)/N\right]^{1/2} \approx 0.044$. For the four spectra (periodograms), the combined power statistic, the minimum power statistic, and the joint power statistic were calculated at the signal frequency. This procedure was repeated 1000 times.

The results are summarized in Figure 5. Panel (a) shows the distribution of power at the signal frequency for the 4000 individual spectra (solid histogram). The spectra were constructed so that the mean of this distribution is $\overline{S}_0 = 5$, and this value is indicated by the dotted vertical line. The solid curve is the expected distribution $e^{-S}$ of noise, i.e. the distribution of power at frequencies other than the signal frequency, and this distribution is reproduced in each panel in the figure. Panel (b) shows the distribution of the value of the combined power statistic at the signal frequency (solid histogram). The mean of the observed distribution (indicated by the vertical dotted line) is about 12.7. Panel (c) shows the distribution of the minimum power statistic at the signal frequency (solid histogram); the mean of this distribution (dotted line) is about 9.1. Finally panel (d) shows the distribution of the joint power statistic at the signal frequency (solid histogram); the mean of this distribution (dotted line) is about 13.0. This figure illustrates how the three statistics lift the power at the signal frequency out of the noise, on average. In running these simulations, we have also verified that (at frequencies other than the signal frequency) the statistics are indeed distributed exponentially.

## 6. DISCUSSION

In preceding sections, we have introduced statistics formed from two or more power spectra that are distributed in the same way as simple power spectra formed from a time series of normally distributed random variables with unit standard deviation. These statistics lead to estimates of the significance level of a peak found at a specified frequency. However, in practice one will be searching for evidence of a peak in a specified band of frequencies. In our neutrino research, for instance, we may look for evidence for a peak in a band corresponding to the range of synodic rotation rates of the solar convection zone or the solar radiative zone. The standard technique for addressing this problem is to estimate the "false-alarm" probability (Press et al. 1992). We estimate (empirically or theoretically) the number N of independent frequencies in this band. Then the probability of obtaining a power S or larger by chance in this band is then given by

$$P(\text{false alarm}) = 1 - \left(1 - e^{-S}\right)^N . \qquad (6.1)$$

Since the statistics introduced in Sections 2, 3 and 4 have exponential distributions, we may apply this formula to these statistics also. If this probability is small, it may be represented approximately by a "corrected" statistic (corrected to take account of the search band) given by

$$G^* \approx G - \ln(N) , \qquad (6.2)$$

with similar expressions for the statistics K and J.

As we pointed out in the introduction, one must be careful in applying the techniques of this article to combine or compare two or more peaks in the same power spectrum since the powers of these peaks may be correlated due to aliasing or for some other reason. Hence the type of analysis described in this article should be regarded as indicative rather than conclusive. If a test indicates that two or more variables may be related, one may proceed to adopt a more robust procedure for significance estimation. For instance, if one is studying a time series that is supposed (on the null hypothesis) to be structure-less, the "shuffle test" can be very useful. (See, for example, Bahcall & Press 1991, Sturrock & Weber 2002, and Walther 1999.) It is interesting to note that since the shuffle test examines only the order of the actual measurement (of the power, or of a combination of powers) and of comparable measurements derived from simulations, normalization of the measurement makes no difference.

One may also look for a specific hypothesis to test (preferably from a different dataset). Then, once a specific hypothesis concerning time series has been specified, it may be evaluated by Bayesian techniques. We plan to investigate this further and report on our findings in a later article.

Bai (2003) has recently reviewed evidence that solar flares tend to exhibit periodicities with periods that are integer (2, 3,…) multiples of a "fundamental period" of approximately 25.5 days. The significance of this result could be assessed by means of a statistic formed from $S(\nu/2)$, $S(\nu/3)$, etc. Wolff (2002) has recently claimed to find somewhat similar patterns in power spectra formed from measurements of the solar radio flux. The statistics we have here introduced should be helpful in the further evaluation of these claims also. We have recently used these statistics in the analysis of Homestake and GALLEX-GNO solar neutrino data to seek and evaluate evidence of r-mode oscillations (see, for instance, Saio 1982) in these data (Sturrock 2003).

We are grateful to Paul Switzer for helpful discussions and to Steve Harris for help with the analytical calculation of Section 4. This research was supported by NSF grant ATM-0097128.

APPENDIX A

We are interested in evaluating the significance of two or more measurements of variables that, if random, have known distributions. When the distributions are normal, one can adopt the chi-square statistic for this purpose. Since we are interested in different distributions, we first recall how it is possible to introduce similar statistics for arbitrary distributions.

We consider variables $x_1$, $x_2$, …, and suppose that these have arbitrary distributions $P_1(x_1)dx_1$, etc., on the relevant null hypotheses. Then the statistic f, defined by

$$f = F(x_1, x_2, ...),  \tag{A.1}$$

is distributed according to the probability distribution function $P_F(f)$, given by the probability integral transformation

$$P_F(f)df = \left[ \iint ... dx_1 dx_2 ... P_1(x_1) P_2(x_2) ... \delta(f - F(x_1, x_2, ...)) \right] df.  \tag{A.2}$$

If each variable has a normal distribution with mean zero and standard deviation unity, and if we adopt the function

$$F(x_1, x_2, ...) = x_1^2 + x_2^2 + ...,  \tag{A.3}$$

f has the familiar chi-square distribution (Bartoszynki & Niewiadomska-Bugaj 1996, p. 757; Rice 1988, p. 168).

If we apply the operation (1.3) to the statistic defined by equation (A.2),

$$\Gamma = -\ln \left( \int_f^\infty df\, P_F(f) \right),  \tag{A.4}$$

we arrive at a statistic that is distributed exponentially. Note that this procedure leads to a unique functional form for $\Gamma$ for any given function $F(x_1, x_2, ...)$.

## APPENDIX B

If

$$Z = S_1 + S_2 + ... + S_n,  \tag{B.1}$$

where $S_1, ..., S_n$ are independent variables, each satisfying an exponential distribution as in (1.1), it is well known that Z satisfies the gamma distribution:

$$P_{C,n}(Z) = \frac{1}{(n-1)!} Z^{n-1} e^{-Z}, \quad (B.2)$$

(Bickel & Doksum 1977, p. 13; Bartoszynki & Niewiadomska-Bugaj 1996, p. 384; Rice 1988, p. 168). We may now carry out the operation of equation (A.4) in order to obtain a statistic that is distributed exponentially. We write

$$G_n(Z) = -\ln\left(\int_Z^\infty dz\, P_{C,n}(z)\right), \quad (B.3)$$

and so arrive at the following formula for the "combined power statistic:"

$$G_n(Z) = Z - \ln\left(1 + Z + \tfrac{1}{2} Z^2 + \ldots + \tfrac{1}{(n-1)!} Z^{n-1}\right). \quad (B.4)$$

APPENDIX C

If we introduce the notation

$$Y = S_1 S_2, \quad (C.1)$$

we see from (A.2) that the probability distribution function for Y is given by

$$P_{J,2}(Y)dY = \int_0^\infty \int_0^\infty du\, dv\, e^{-u-v} \delta(Y - uv) dY, \quad (C.2)$$

i.e., by

$$P_{J,2}(Y) = \int_0^\infty \frac{du}{u} e^{-u - Y/u}. \quad (C.3)$$

We may use equation (A.4) to transform this statistic to one, that we call the "joint power statistic," that is distributed exponentially:

$$J_2(Y) = -\ln\left(\int_Y^\infty dy\, P_{J,2}(y)\right), \tag{C.4}$$

which becomes

$$J_2 = -\ln\left(\int_0^\infty du\, e^{-u-Y/u}\right). \tag{C.5}$$

This is found to be expressible as

$$J_2 = -\ln\left(2Y^{1/2} K_1(2Y^{1/2})\right), \tag{C.6}$$

where $K_1$ is the Bessel function of the second kind.

We see that it is more convenient to introduce the geometric mean

$$X = (S_1 S_2)^{1/2}, \tag{C.7}$$

so that (C.6) becomes

$$J_2 = -\ln(2X K_1(2X)). \tag{C.8}$$

The leading terms of the asymptotic expression for $J_2$ are as follows:

$$J_2 \to 2X - \tfrac{1}{2}\ln(\pi) - \tfrac{1}{2}\ln(X) \text{ as } X \to \infty. \tag{C.9}$$

## APPENDIX D

We now consider joint power statistics of higher orders. If

$$Y = S_1...S_n , \tag{D.1}$$

we see from (A.2) that the probability distribution function for $Y$ is given by

$$P_{J,n}(Y)dY = \left[\int_0^\infty dx_1...\int_0^\infty dx_n e^{-x_1-...-x_n} \delta(Y - x_1...x_n)\right]dY . \tag{D.2}$$

By examining the corresponding formula for $P_{n+1}(Y)$, we find that

$$P_{J,n+1}(Y) = \int_0^\infty dx e^{-x} P_{J,n}(Y/x) . \tag{D.3}$$

In terms of the cumulative distribution functions,

$$C_{J,n}(Y) = \int_Y^\infty dy P_{J,n}(y) , \tag{D.4}$$

the iterative relation (D.3) becomes

$$C_{J,n+1}(Y) = \int_0^\infty dx e^{-x} C_{J,n}(Y/x) . \tag{D.5}$$

We have used this iterative relation to calculate joint power statistics of the third and fourth orders. Application of the iterative relation (D.5) gives the following expressions for $C_{J,3}$ and $C_{J,4}$:-

$$C_{J,3}(X) = \int_0^\infty dx e^{-x} 2\frac{X^{3/2}}{x^{1/2}} K_1\left[2X^{3/2}/x^{1/2}\right] \tag{D.6}$$

and

$$C_{J,4}(X) = \int_0^\infty dx \int_0^\infty dx' e^{-(x+x')} \frac{2X^2}{(xx')^{1/2}} K_1\left[2X^2/(xx')^{1/2}\right] . \tag{D.7}$$

The joint power statistics of third and fourth orders are then given by

$$J_3(X) = -\ln[C_{J,3}(X)], \quad J_4(X) = -\ln[C_{J,4}(X)]. \tag{D.8}$$

## APPENDIX E

We now develop approximate formulae for the joint power statistics of second, third and fourth orders. After trial and error, we were led to consider the form

$$J = \frac{aX^2}{b+X}. \tag{E.1}$$

We could have attempted a nonlinear process for determining the optimum values of the coefficients a and b. However, it was simpler and adequate to rewrite (E.1) as

$$X^2 a - Jb = XJ. \tag{E.2}$$

Given a set of values X and J, determined analytically for J2 or by computation for J3 and J4, we could then determine the coefficients a and b by the least-squares process. The results of these calculations are given in Section 4.

## APPENDIX F

Note that one may obtain approximate expressions for the higher-order joint power statistics by combining the results of Appendix D and Appendix E. Beginning with an approximate formula for $J_2, J_3, \text{or } J_4$, as developed in Appendix E, we may compute the cumulative distribution function by using (1.2). We may now use (D.5) to obtain an estimate of the cumulative distribution function for the statistic of next higher order, from which we may obtain the statistic itself by means of (1.3). We may then seek an approximate formula for this statistic by the procedure of Appendix E. This procedure may be repeated for as long as the approximations prove to be acceptable.

FIGURE CAPTIONS

Figure 1. Combined power statistics of second, third and fourth orders: G2 ('o'), G3 ('x'), and G4 ('+').

Figure 2. Joint power statistic of second, third and fourth orders, J2 ('o'), J3 ('x'), and J4 ('+'). The solid line shows estimates from the approximate expressions (4.3), (4.5), and (4.6). For J2, the points ('o') are calculated from the analytical expression (4.2). For J3 ('x') and J4 ('+'), the points are calculated by the method outlined in Appendix D.

Figure 3. Four synthetic spectra, each with a signal of power 5 at $\nu = 20$.

Figure 4. The combined power statistic, minimum power statistic, and joint power statistic, formed from the four synthetic spectra shown in Figure 3.

Figure 5. Distribution of statistics evaluated at the signal frequency for 1000 sets of four spectra produced from synthetic time series: (a) power spectrum; (b) combined power statistic; (c) minimum power statistic; and (d) joint power statistic. The exponential curve in each panel is the distribution of power at frequencies other than the signal frequency. For each panel, the vertical line indicates the mean of the computed distribution at the signal frequency.

FIGURES

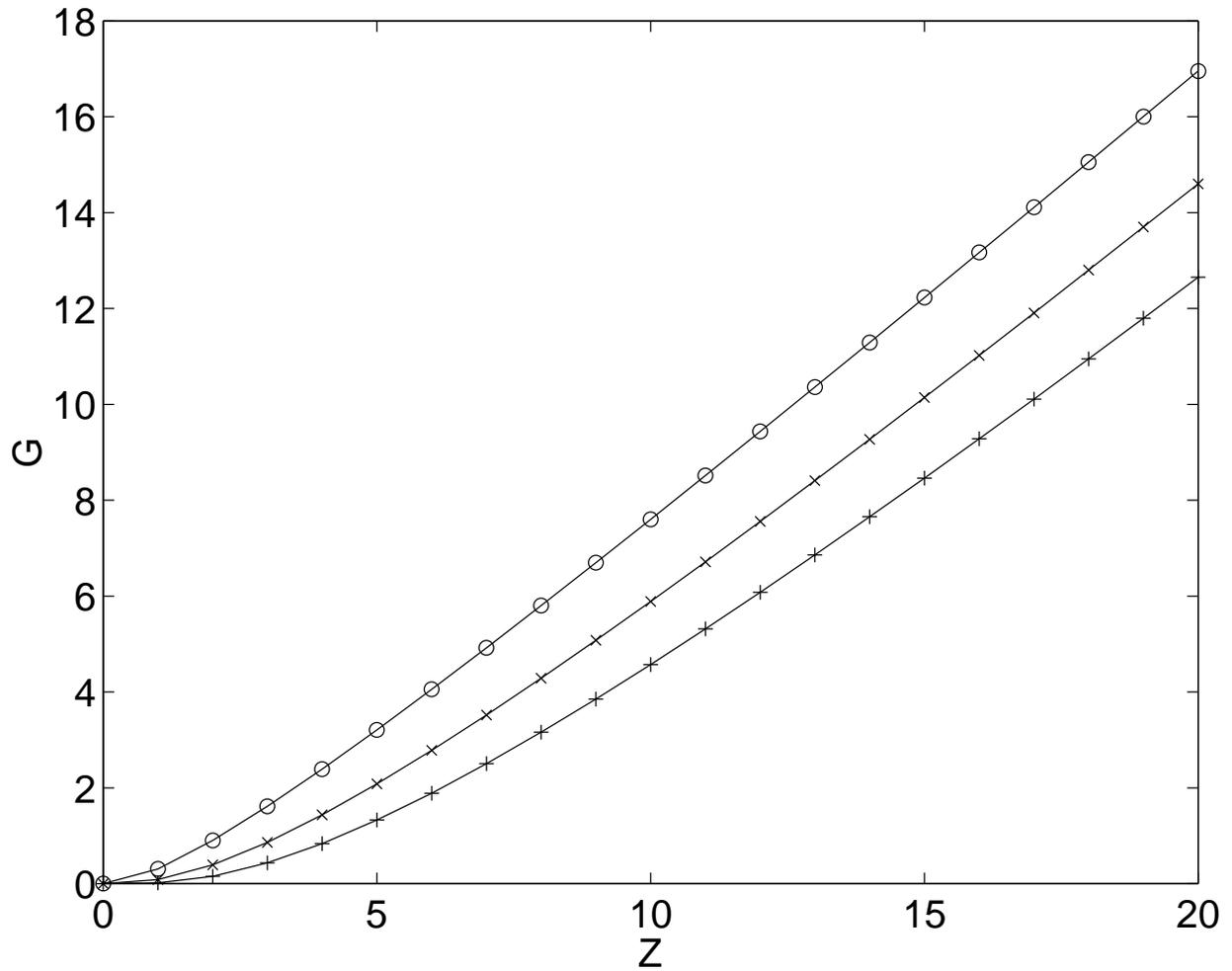

Figure 1. Combined power statistics of second, third and fourth order: G2 ('o'), G3 ('x'), and G4 ('+').

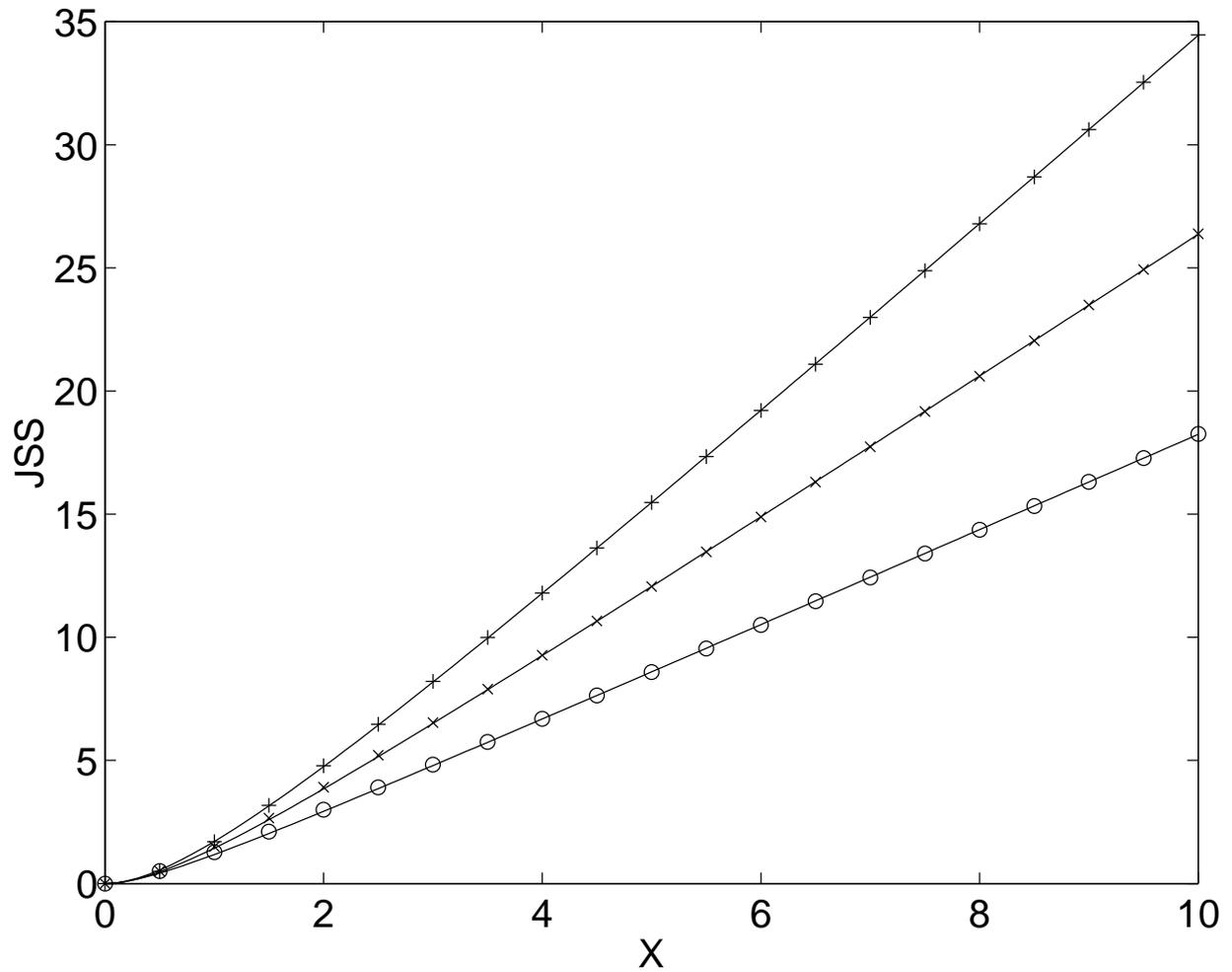

Figure 2. Joint power statistic of second, third and fourth order, J2 ('o'), J3 ('x'), and J4 ('+'). The solid line shows estimates from the approximate expressions (4.3), (4.5), and (4.6). For J2, the points ('o') are calculated from the analytical expression (4.2). For J3 ('x'), and J4 ('+'), the points are calculated by the method outlined in Appendix D.

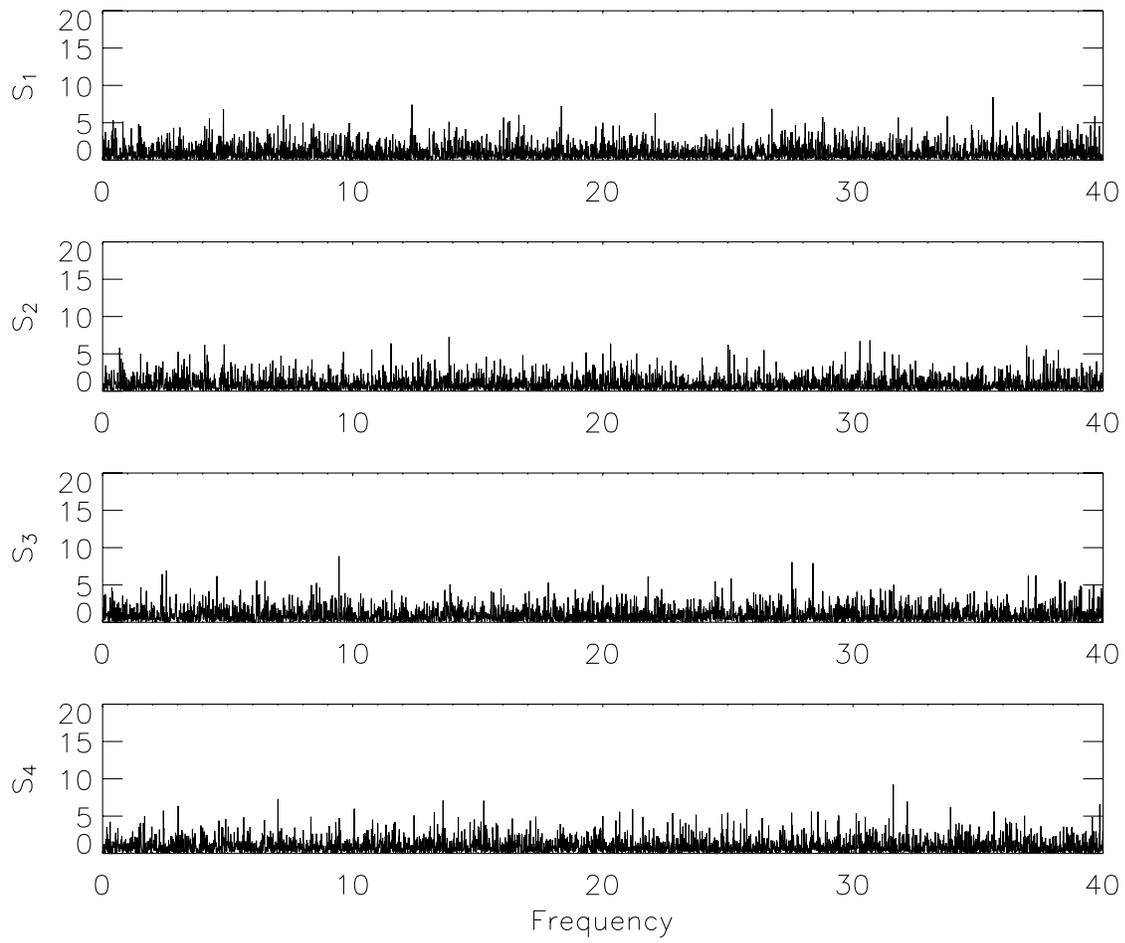

Figure 3. Four synthetic spectra, each with a signal of power 5 at $\nu = 20$.

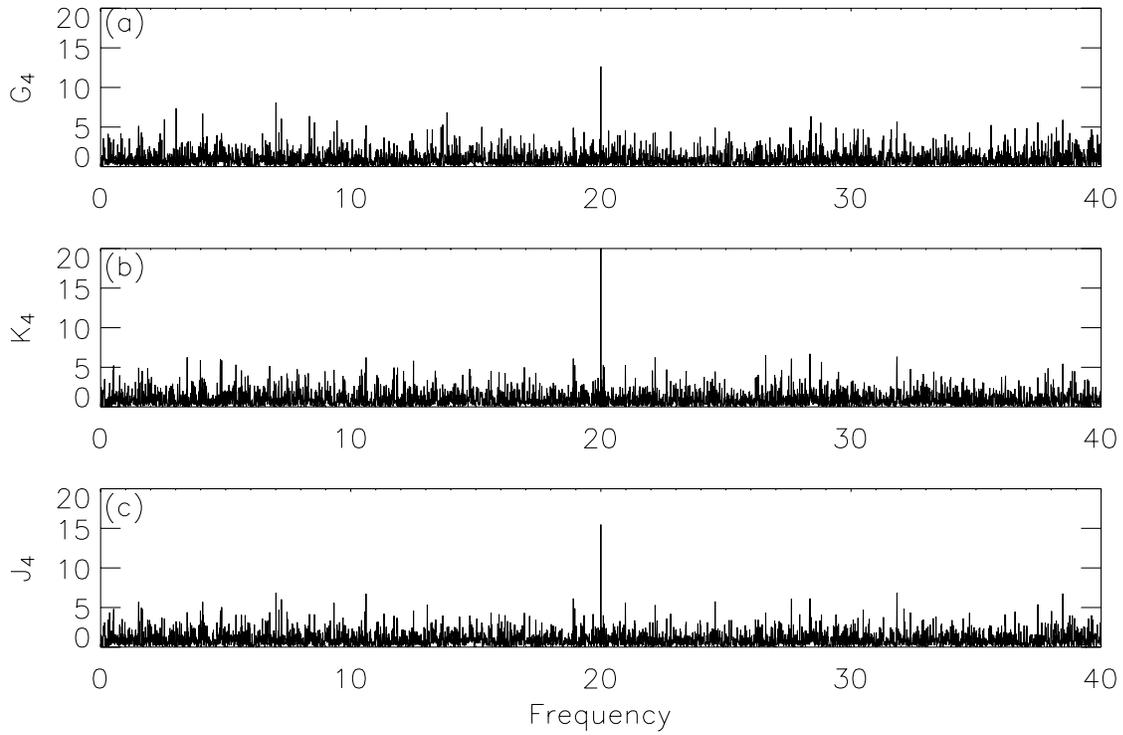

Figure 4. The combined power statistic, minimum power statistic, and joint power statistic, formed from the four synthetic spectra shown in Figure 3.

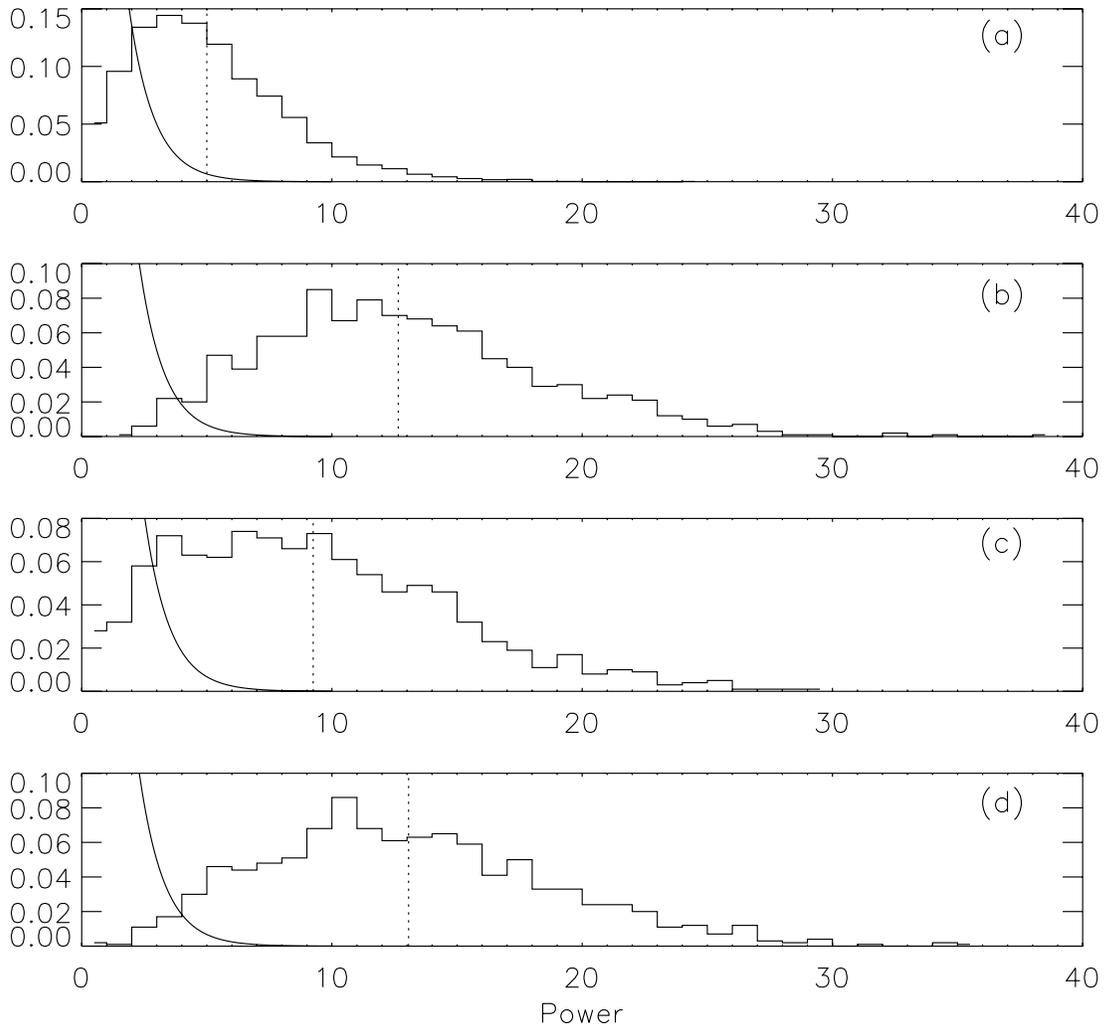

Figure 5. Distribution of statistics evaluated at the signal frequency for 1000 sets of four spectra produced from synthetic time series: (a) power spectrum; (b) combined power statistic; (c) minimum power statistic; and (d) joint power statistic. The exponential curve in each panel is the distribution of power at frequencies other than the signal frequency. For each panel, the vertical line indicates the mean of the computed distribution at the signal frequency.

26